\documentclass{webofc}
\usepackage[varg]{txfonts}   
\usepackage{physics}
\usepackage{hyperref}
\begin{document}

\title{Particle Track Reconstruction with Quantum Algorithms}

\author{\firstname{Cenk} \lastname{Tüysüz}\inst{1,2}\fnsep\thanks{\email{cenk.tuysuz@cern.ch
        }} 
        \and
        \firstname{Federico} \lastname{Carminati}\inst{3}
        \and
        \firstname{Bilge} \lastname{Demirköz}\inst{1}
        \and
        \firstname{Daniel} \lastname{Dobos}\inst{4,6}
        \and
        \firstname{Fabio} \lastname{Fracas}\inst{3}
        \and
        \firstname{Kristiane} \lastname{Novotny}\inst{4}
        \and
        \firstname{Karolos} \lastname{Potamianos}\inst{4,5}
        \and
        \firstname{Sofia} \lastname{Vallecorsa}\inst{3}
        \and
        \firstname{Jean-Roch} \lastname{Vlimant}\inst{7}
}

\institute{
        Middle East Technical University, Ankara, Turkey
\and
        STB Research, Ankara, Turkey
\and
        CERN, Geneva, Switzerland
\and
        gluoNNet, Geneva, Switzerland
\and
        DESY, Hamburg, Germany
\and
        Lancaster University, Lancaster, UK
\and
        California Institute of Technology, Pasadena, California, USA
          }

\abstract{
Accurate determination of particle track reconstruction parameters will be a major challenge for the High Luminosity Large Hadron Collider (HL-LHC) experiments. The expected increase in the number of simultaneous collisions at the HL-LHC and the resulting high detector occupancy will make track reconstruction algorithms extremely demanding in terms of time and computing resources. The increase in number of hits will increase the complexity of track reconstruction algorithms. In addition, the ambiguity in assigning hits to particle tracks will be increased due to the finite resolution of the detector and the physical “closeness” of the hits. Thus, the reconstruction of charged particle tracks will be a major challenge to the correct interpretation of the HL-LHC data. Most methods currently in use are based on Kalman filters which are shown to be robust and to provide good physics performance. However, they are expected to scale worse than quadratically. Designing an algorithm capable of reducing the combinatorial background at the hit level, would provide a much “cleaner” initial seed to the Kalman filter, strongly reducing the total processing time. One of the salient features of Quantum Computers is the ability to evaluate a very large number of states simultaneously, making them an ideal instrument for searches in a large parameter space. In fact, different R\&D initiatives are exploring how Quantum Tracking Algorithms could leverage such capabilities. In this paper, we present our work on the implementation of a quantum-based track finding algorithm aimed at reducing combinatorial background during the initial seeding stage. We use the publicly available dataset designed for the kaggle TrackML challenge.
}
\maketitle
\section{Introduction}
\label{sec-1}

Latest developments in Quantum Computing, significantly reduced the time needed to resolve certain problems"\cite{google}. This resulted in a search for new methods to boost current algorithms whose computational complexity depend on the size of the dataset worse than polynomial. 

The upcoming upgrade of Large Hadron Collider (LHC) at CERN to High Luminosity (HL-LHC) will increase the number of collisions. The HL-LHC upgrade will bring many challenges. Particle track reconstruction is one of the challenges \cite{ref-hilumi}. Current algorithms have trouble scaling up to higher collision rates. Therefore, researchers are trying new methods to tackle the problem such as the use of Graph Neural Networks\cite{HepTrkX} and Quantum Computing\cite{qalg1,qalg2,qalg3}.

When a particle passes through a tracking detector layer, a signal called a "hit" is generated. The dataset contains precise locations of these hits and their particle identifications as labels. The challenge is to associate hits that are belonging to the same initial particle/track.

The HepTrkX team proposed a Graph Neural Network implementation for particle track reconstruction that uses the kaggle TrackML challenge dataset\cite{HepTrkX,TrackML}. The simulated dataset and the challenge was created by CERN scientists to invite machine learning experts to come up with novel methods to track reconstruction. Even though the Kaggle competition is concluded, the dataset is still being actively used by many researchers in the field of track reconstruction to benchmark their results.

The speed-up provided by Quantum Algorithms may play an important role in the future of track reconstruction in particle physics experiments. In this work, we present an exploratory look at the HepTrkX\cite{heptrkx_github} project from a Quantum Computing perspective to evaluate the capabilities of Quantum Computing along with Deep Learning algorithms \cite{HepTrkX}.

\section{The Dataset and Classical Approach}
\label{sec-2}

\begin{figure}[h]
\centering
\includegraphics[width=13cm]{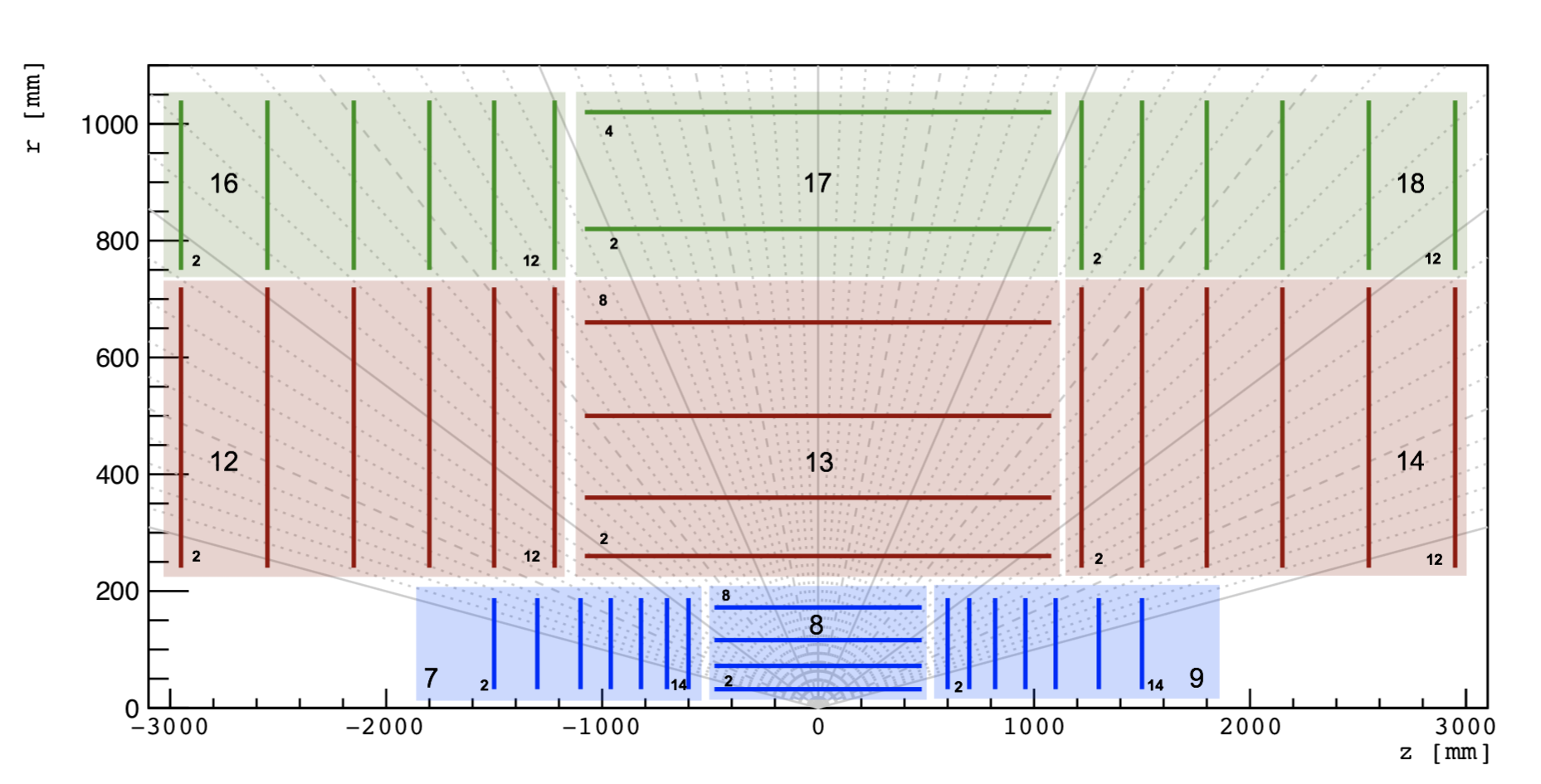}
\caption{TrackML Detector Layout \cite{TrackML}.}
\label{fig-detector}       
\end{figure}

The TrackML dataset consists of simulated measurements of many detector layers. The detector layers are arranged using a model layout that is common to most LHC experiments. In this layout there are several detectors that span a cylindrical geometry. All the detectors are rotated such that their center sees the collision center. In this geometry, particle beams collide on the $z$ axis around $z=0$ which is the center of collisions and also the center of the detector. The layout of the detectors can be seen in Figure~\ref{fig-detector}.

The complex layout of detectors allow better precision. However, they also increase the computational complexity as the particles might pass through both horizontal (barrel) and vertical (endcap) layers of detectors. For simplicity, the model only uses the barrel region of the detector. Therefore, the dataset contains only the layers in regions 8, 13 and 17.

The work presented in this paper uses the same preprocessing steps as those made by the HepTrkX team. This step is used to create an initial graph and labels from the TrackML dataset which contain particle momentum and spatial coordinates of detector measurements which are called "hits" and particle identification numbers. 

The initial graph is created by connecting all logically possible combinations of hits and then applying loose selection criteria in order to decrease connectivity of the initial graph. This selection prevent connections between far detectors and acute angles which are physically not possible. The cuts used are given in Table~\ref{table-cuts}. 

\begin{table}[h]
\centering
\caption{Cuts applied to TrackML dataset for preprocessing.}
\label{table-cuts}     
\begin{tabular}{|c|c|}
\hline
$\left|p_T\right|$ & $>1 GeV$     \\\hline
$\Delta\phi$ & $<0.0006$  \\\hline
$z_0$ & $<100 mm$  \\\hline
$\eta$ & $[-5,5]$  \\\hline
\end{tabular}
\end{table}

The coordinate definitions are as follows. $\phi$ is the angle along the transverse plane ($xy$ plane)  and $\left|p_T\right|$ is the magnitude of momentum along the same plane. $\eta$ is the psuedorapidity which is a parameter widely used in particle physics as a measure of the azimuthal angle with respect to the beam axis.

The graphs created using the TrackML dataset are further divided into 8 in $\phi$ direction and into 2 in $z$ direction to reduce the size of the graphs for a single event. A single event contains $\sim8k$ hits requiring tremendous amounts of memory. By dividing a graph in its symmetry axes to 16, the same operations can be applied with less memory. This becomes handy when using limited memory systems to train the model.

\begin{figure}[h]
\centering
\includegraphics[width=13cm]{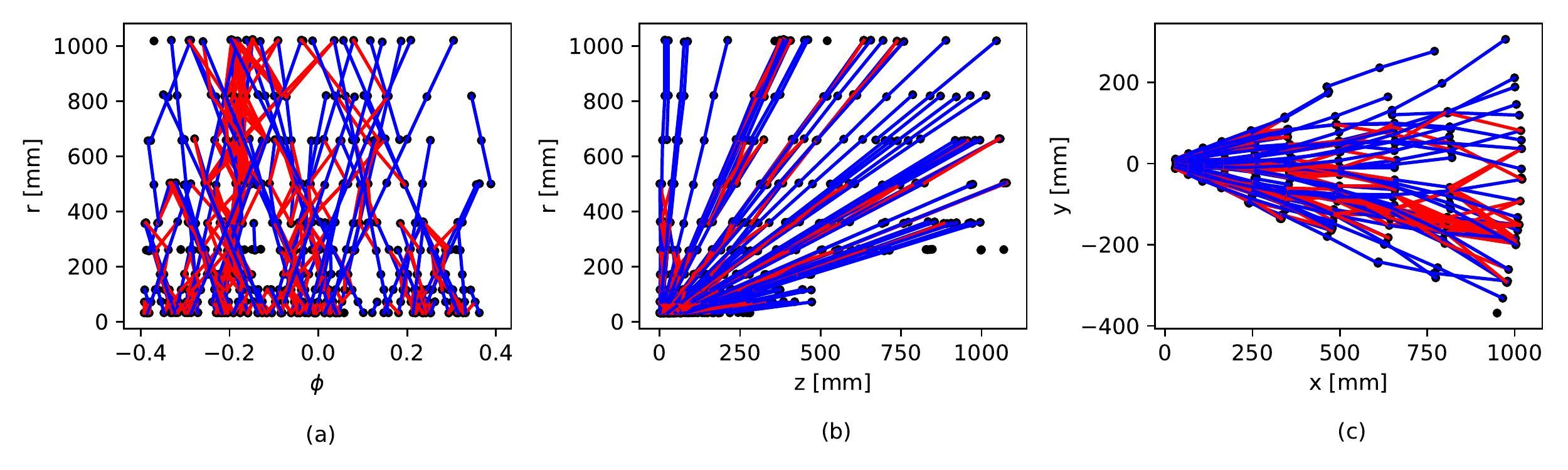}
\caption{1 of 16 subgraphs created from a single event. (a,b) are subgraphs in cylindrical coordinates and (c) is a subgraph in Cartesian coordinates. Red represents Ground Truth, while Blue shows Fake edges created using loose cuts.}
\label{fig-graph}       
\end{figure}

\begin{figure}
\centering
\includegraphics[width=13cm]{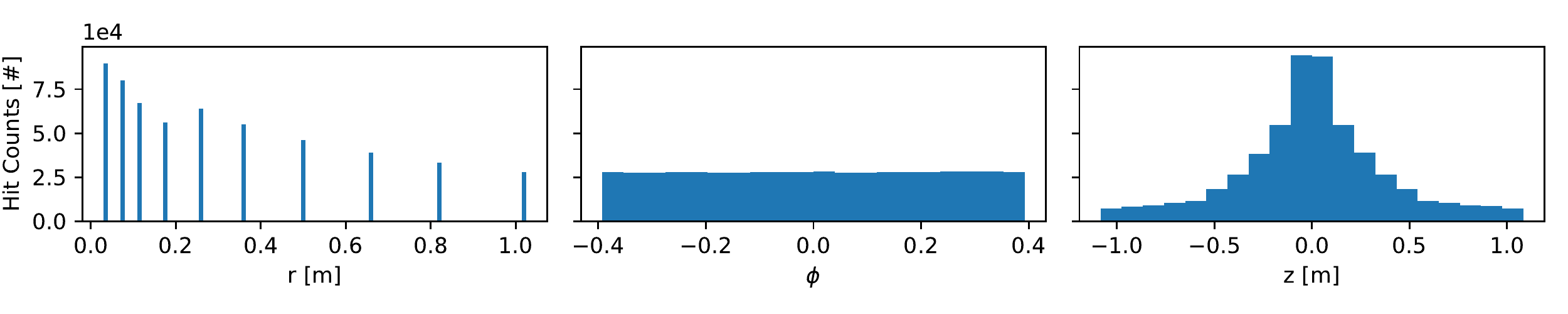}
\caption{Histogram of hits crated using 1600 subgraphs in cylindrical coordinates. The distribution of r and z can be explained by referring to the geometry of the detector in Figure~\ref{fig-detector}. The distribution in $\phi$ is uniform as expected since the geometry is symmetric along the transverse  plane.}
\label{fig-data}       
\end{figure}

As a first step, 100 events from the TrackML dataset are used to create 1600 subgraphs using the selection defined in Table~\ref{table-cuts}. The reprocessed dataset is created using $1.0\%$ of the TrackML dataset. The size of the dataset is kept intentionally small to reduce simulation times in order to speed up prototyping. Figure~\ref{fig-graph} shows an initial subgraph after preprocessing. Histograms showing the distribution of hits from the preprocessed data in cylindrical coordinates are given in Figure~\ref{fig-data}.

The HepTrkX team proposed a GNN (Graph Neural Network) to perform segment classification. The model consists of 3 types of networks. The first one is an Input Network which takes a graph and maps it to higher dimensions. The second one is an Edge Network which takes node information from all edges in a graph and computes the edge information. The last network is a Node Network which computes hidden node features by looking at neighbouring nodes of each node. Edge and Node Networks are applied recursively after the execution of the Input Network. The model which is tested using the same TrackML dataset showed excellent performance. The model scores are $99.5\%$ purity, $98.7\%$ efficiency, and overall accuracy of $99.5\%$ with 0.5 threshold \cite{HepTrkX}. Readers can refer to \cite{ref-tensorflow} for definitions of metrics.

\section{Quantum Circuits as Neural Networks}
\label{sec-3}

The increase in computational power in the last two decades allowed scientists to deploy efficient machine learning models. Neural Networks use the immense power of classical computing to represent hidden features in data using millions of artificial neurons. However, even with today's computational power some tasks still take a considerable amount of time as the complexity of the tasks increases.

Quantum computing allows using entanglement, enabling the introduction of correlations that are not classically available to the neural network models. Additionally, the models can be extended to higher dimensions much faster as the dimension of the Hilbert Space is $2^n$, with n being the number of qubits.

Quantum circuits have been previously shown to perform binary classification. Many Quantum circuit models in literature were considered \cite{ttn,mps}. In this work, hierarchical quantum circuits have been selected to replace Neural Network layers, due to their high accuracy and robustness against noise \cite{ttn}. 

The constructed models are implemented using Tensorflow~\cite{ref-tensorflow} and Pennylane~\cite{ref-pennylane}. Pennylane is an automatic differentiation tool for quantum circuits and is used to calculate the gradients \cite{ref-pennylane}. Readers can also refer to \cite{farhi-classification} for a more theoretical view.

\section{Quantum Computing Integration} 
\label{sec-4}

\begin{figure}[h]
\centering
\includegraphics[width=13cm]{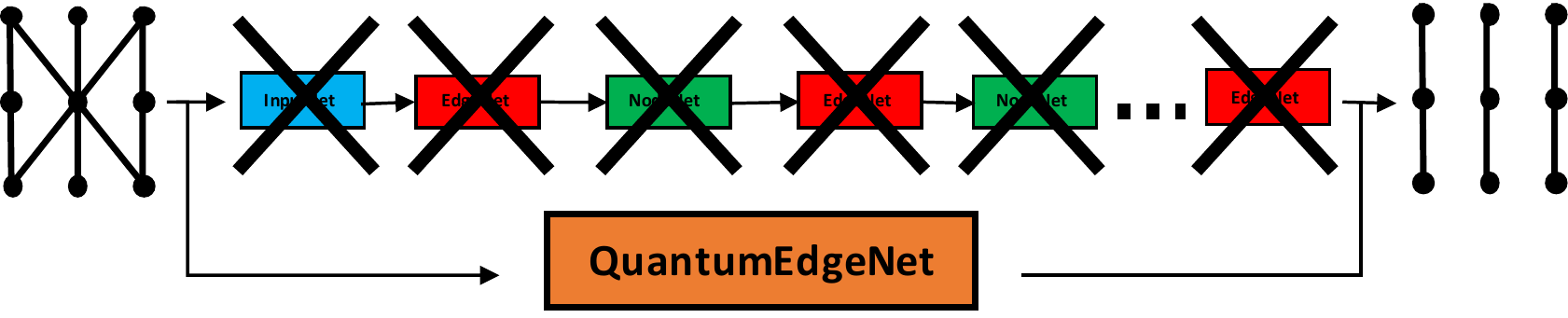}
\caption{The HepTrkX GNN structure and the QuantumEdgeNet model used in this work.}
\label{fig-network}       
\end{figure}

Transforming a well-performing Graph Neural Network to a Quantum Computing structure requires many modifications. To go step by step, this work only replaces the Edge Network of HepTrkX and does not use the Input and Node Network for simplicity. Therefore, the network only takes spatial information of nodes and computes the probability of an edge being true or fake. The original HepTrkX GNN model and the model in this work can be seen in Figure~\ref{fig-network}.

The Tree Tensor Network (TTN) is chosen among the hierarchical quantum classifiers as the quantum circuit to replace the neural network layer. The first attempt is made with no hidden dimensions to test the capabilities of the structure. In later work, hidden dimensions will be included.

The Quantum Edge Network is applied to all edges one by one. All edges contain 2 node information which consists of 3 spatial coordinates each summing up to 6 data points for every edge. The coordinate information is encoded in 6 qubits, mapping each one to $0-2\pi$. This encoding method is chosen for its simplicity as other methods introduce additional computational complexities to use less qubits negating the gain in the number of qubits \cite{Aaronson2015}. The data points are used as inputs to $R_y$ rotation gates to encode the information as the angle between $\ket{0}$ and $\ket{1}$ states.

\begin{equation}
\label{eq1}
R_y(\theta)\ket{0} = \cos(\theta/2)\ket{0} + \sin(\theta/2)\ket{1}
\end{equation}

\begin{figure}[h]
\centering
\includegraphics[width=8cm]{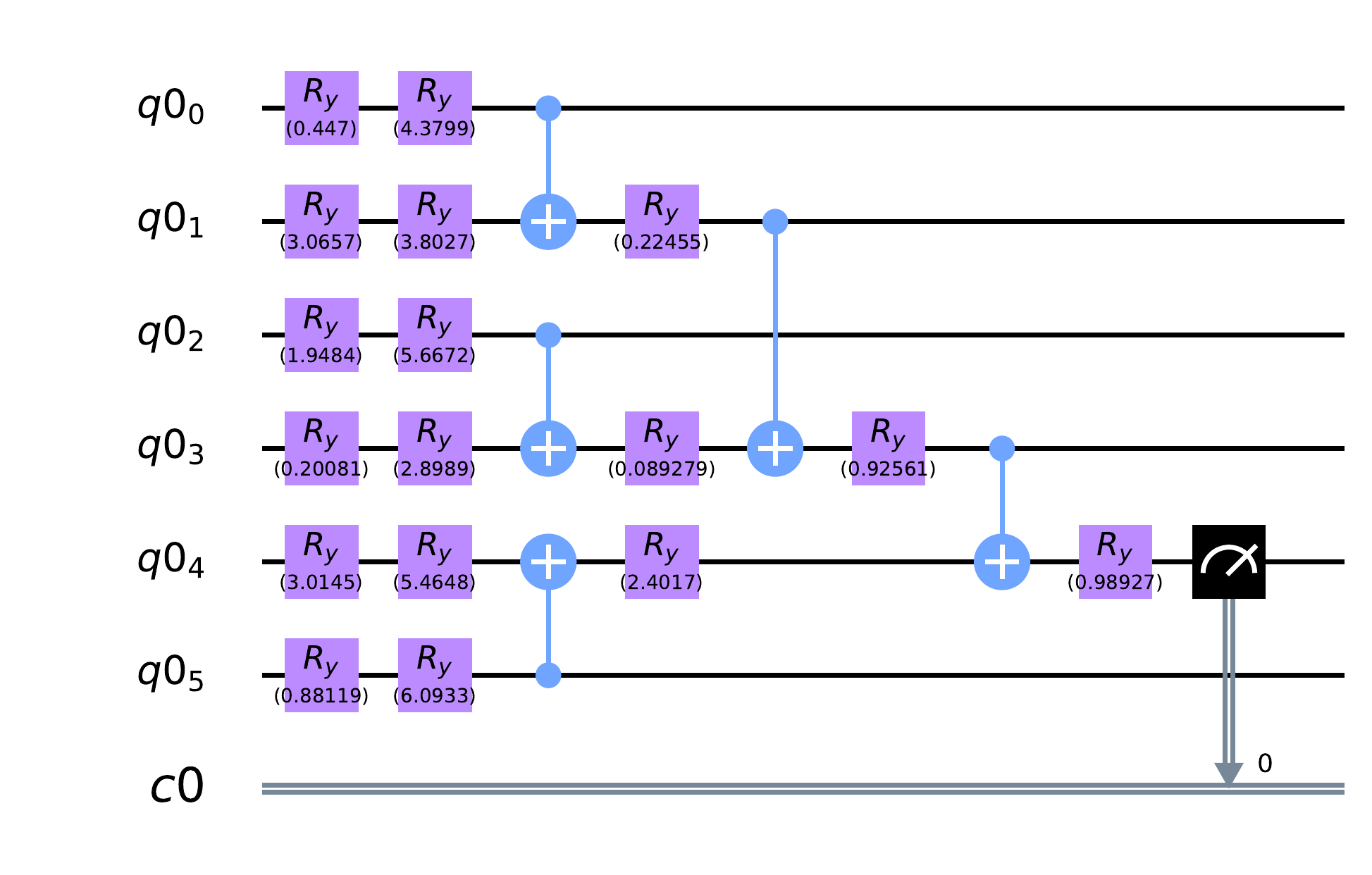}
\caption{6 Qubit Tree Tensor Network (TTN) representation of the Quantum Edge Network.}
\label{fig-circuit}       
\end{figure}

After encoding the input in qubits, the TTN circuit is applied. The TTN circuit contains $R_y$ and $CNOT$ gates. $R_y$ gates starts with random parameters to be tuned later and $R_y$ gates rotate the state according to the parameter's value. The $CNOT$ gate is used to introduce correlation between qubits so that their values are not independent. At the end of the circuit, there is a measurement. The input encoding layer and TTN circuit structure plotted using Qiskit~\cite{ref-qiskit} and matplotlib~\cite{ref-matplotlib} can be seen in Figure~\ref{fig-circuit}.

The quantum circuit is required to be run multiple times as the result of a single measurement is either 0 or 1. Therefore, the circuit is run 1000 times and the average of the outputs is used to determine the probability of an edge being true or fake. The number of circuit runs, in general called \textit{shots}, should be selected carefully and the best value depends on a trade off between error rate and run time.

\begin{figure}[h]
\centering
\includegraphics[width=6cm]{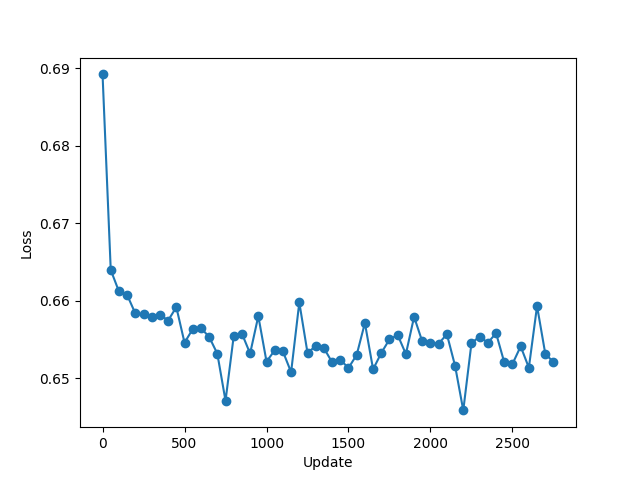}%
\includegraphics[width=6cm]{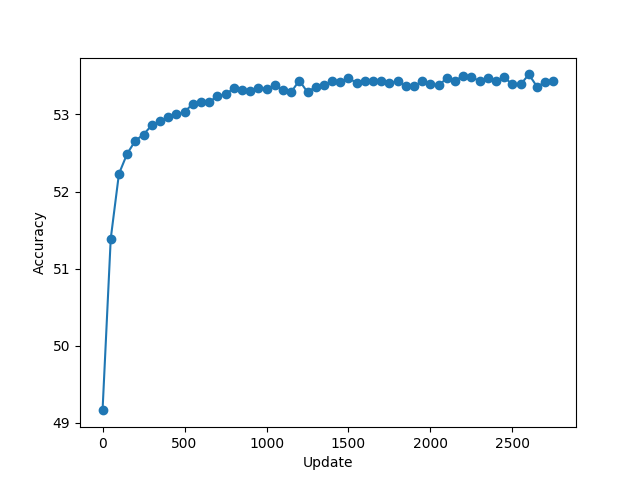}
\caption{Training Loss (on the left) and Validation Accuracy (on the right) of the TTN in 2 full epochs. (1 epoch = 1440 updates)}
\label{fig-results}       
\end{figure}

The network is trained over 2 epochs. The subgraphs are divided randomly into training and test sets with a 9:1 ratio. The model is trained using stochastic gradient descent and weighted binary cross entropy \cite{ref-tensorflow}. Training performance of the model can be seen in Figure~\ref{fig-results}.

The results in Figure~\ref{fig-results} show that the model can learn features of the graph data. The accuracy achieved is considerably small, but this is mainly due to over simplification of the model. The network showcased here is a proof of principle prototype of a complete Quantum Graph Neural Network structure.

\section{Future Work} 
\label{sec-5}

The work presented here is the first step towards a Quantum Graph Neural Network that can classify tracks with high precision and accuracy. The final structure will include; 
\begin{itemize}
    \item The Node Network as another quantum circuits to learn features among neighbouring nodes.
    \item  Recursive iterations of Node and Edge Networks to pass the information through nodes
    \item An input layer to include hidden layers, which do learn in a large Hilbert Space.
\end{itemize}

The up-to-date model and all source codes can be accessed through \cite{ref-heptrkx-quantum}.

\section{Conclusion} 
\label{sec-6}
In this work, we show that it is possible to implement Quantum Graph Neural Network based approaches for the track reconstruction problem. Although, the current model is not complete, it shows promising preliminary results towards a quantum circuit based algorithm that can be run using a universal quantum computer. This work only uses simulations and does not consider practical applicability at the moment.
\section{Acknowledgments} 
\label{sec-7}
Part of this work was conducted at "\textit{iBanks}", the AI GPU cluster at Caltech. We acknowledge NVIDIA, SuperMicro and the Kavli Foundation for their support of "\textit{iBanks}". This work was partially supported by Turkish Atomic Energy Authority (TAEK) (Grant No: 2017TAEKCERN-A5.H6.F2.15). Cenk Tüysüz thanks Oral Okan and Egemen Sert from STB for their valuable discussions.

\bibliography{main.bib}

\begin{thebibliography}{17}

\bibitem{google}
F.~Arute, K.~Arya, R.~Babbush, D.~Bacon, J.C. Bardin, R.~Barends, R.~Biswas,
  S.~Boixo, F.G.S.L. Brandao, D.A. Buell et~al., Nature \textbf{574} (2019),
  \texttt{arXiv:1910.11333}

\bibitem{ref-hilumi}
G.~Apollinari, O.~Bruening, T.~Nakamoto, L.~Rossi, \emph{High luminosity large
  hadron collider hl-lhc} (2017), \texttt{arXiv:1705.08830}

\bibitem{HepTrkX}
S.~Farrell, P.~Calafiura, M.~Mudigonda, Prabhat, D.~Anderson, J.R. Vlimant,
  S.~Zheng, J.~Bendavid, M.~Spiropulu, G.~Cerati et~al. (2018),
  \texttt{arXiv:1810.06111}

\bibitem{qalg1}
I.~Shapoval, P.~Calafiura (2019), \texttt{arXiv:1902.00498}

\bibitem{qalg2}
F.~Bapst, W.~Bhimji, P.~Calafiura, H.~Gray, W.~Lavrijsen, L.~Linder (2019),
  \texttt{arXiv:1902.08324}

\bibitem{qalg3}
A.~Zlokapa, A.~Anand, J.R. Vlimant, J.M. Duarte, J.~Job, D.~Lidar, M.~Spiropulu
  (2019), \texttt{arXiv:1908.04475}

\bibitem{TrackML}
S.~Amrouche, L.~Basara, P.~Calafiura, V.~Estrade, S.~Farrell, D.R. Ferreira,
  L.~Finnie, N.~Finnie, C.~Germain, V.V. Gligorov et~al. (2019),
  \texttt{arXiv:1904.06778}

\bibitem{heptrkx_github}
\urlstyle{tt}\url{{https://heptrkx.github.io/}}

\bibitem{ref-tensorflow}
M.~Abadi, A.~Agarwal, P.~Barham, E.~Brevdo, Z.~Chen, C.~Citro, G.S. Corrado,
  A.~Davis, J.~Dean, M.~Devin et~al., \emph{{TensorFlow}: Large-scale machine
  learning on heterogeneous systems} (2015), software available from
  tensorflow.org, \urlstyle{tt}\url{https://www.tensorflow.org/}

\bibitem{ttn}
E.~Grant, M.~Benedetti, S.~Cao, A.~Hallam, J.~Lockhart, V.~Stojevic, A.G.
  Green, S.~Severini, npj Quantum Information \textbf{4}, 17 (2018)

\bibitem{mps}
A.S. Bhatia, M.K. Saggi, A.~Kumar, S.~Jain, Neural Computation \textbf{31},
  1499 (2019), \texttt{arXiv:1905.01426v1}

\bibitem{ref-pennylane}
V.~Bergholm, J.~Izaac, M.~Schuld, C.~Gogolin, C.~Blank, K.~McKiernan,
  N.~Killoran, pp. 1--12 (2018), \texttt{arXiv:1811.04968}

\bibitem{farhi-classification}
E.~Farhi, H.~Neven (2018), \texttt{arXiv:1802.06002}

\bibitem{Aaronson2015}
S.~Aaronson, Nat. Phys. \textbf{11}, 291 (2015)

\bibitem{ref-qiskit}
H.~Abraham, I.Y. Akhalwaya, G.~Aleksandrowicz, T.~Alexander, G.~Alexandrowics,
  E.~Arbel, A.~Asfaw, C.~Azaustre, AzizNgoueya, P.~Barkoutsos et~al.,
  \emph{Qiskit: An open-source framework for quantum computing} (2019)

\bibitem{ref-matplotlib}
J.D. Hunter, Computing in Science \& Engineering \textbf{9}, 90 (2007)

\bibitem{ref-heptrkx-quantum}
\urlstyle{tt}\url{{https://github.com/cnktysz/HepTrkX-quantum/}}

\end{thebibliography}

\end{document}